\documentclass[]{aastex631}
\usepackage{hhline,colortbl}

\usepackage{svg}
\usepackage{amssymb}
\begin{document}
\newcommand{\gro}{GRO\,J1655-40}

\title
{Insights into density and location diagnostics of photo-ionized outflows in X-ray binaries

}

\correspondingauthor{Sharon Mitrani}
\email{sharonm@campus.technion.ac.il}

\author{Sharon Mitrani}
\affiliation{Department of Physics, Technion, Haifa, Israel \\ \\}

\author{Ehud Behar}
\affiliation{Department of Physics, Technion, Haifa, Israel \\ \\}







\begin{abstract}
The population of meta-stable levels is key to high precision density diagnostics of astrophysical plasmas. 
In photo-ionized plasmas, density is used to infer the distance from the ionizing source, which is otherwise difficult to obtain. Perfecting models that compute these populations is thus 
crucial.
The present paper presents a semi-analytic hydrogenic approximation for assessing the relative importance of different processes in populating atomic levels. This approximation shows that in the presence of a radiation source, photo- and collisional- excitations are both important over a wide range of plasma temperatures and ionizing spectra, while radiative recombination is orders of magnitude weaker. 
The interesting case of Fe$^{+21}$ with a collisional radiative model with photo-excitation demonstrates this effect. 
The population of the first excited meta-stable level in Fe$^{+21}$ is sensitive to the electron number density in the critical range of $n_e=10^{12}-10^{15}\,\rm{cm}^{-3}$; it was observed to be significantly populated in the X-ray spectrum of the 2005 outburst of the X-ray binary \gro. 
The present model shows that photo-excitation is the predominant process indirectly populating the meta-stable level. 
For the photo-ionized plasma in the \gro\ outflow, 
the model indicates a measured value of $n_e=(2.6 \pm 0.5)\times10^{13}\,\rm{cm}^{-3}$ implying a distance from the source of $r=(4.4 \pm 0.4)\times10^{10}$\,cm.
Finally, we show how the computed critical density and distance of Fe$^{+21}$ yield the correct ionization parameter of the ion, independent of ionization balance calculations.
\end{abstract}




\section{Introduction} \label{sec:intro}
Emission and absorption from meta-stable levels is a powerful diagnostic of the electron number density in highly ionized plasmas. At low densities, excitation from the ground level dominates the population of excited levels, and line ratios are independent of the density. 
At higher densities, electron impact transitions between excited levels significantly change the relative population of meta-stable levels. Once this effect becomes dominant, the line spectrum starts depending on the electron density, until at the highest densities, all level populations tend to the Boltzmann distribution.

He-like X-ray lines from various elements between C and Si were first used by \citet{Gabriel1969} to infer electron densities of $n_e = 10^{8} - 10^{12}$\,cm$^{-3}$ in the hot solar corona. The He-like line diagnostics can be generalized to low temperature photo-ionized plasmas 
 \citep{Porquet2000}, 
e.g. as found in Active Galactic Nuclei (AGNs) and X-Ray Binaries (XRBs).
In these cases, photo-excitation (PE) processes also populate excited levels through resonant transitions \citep{Sako2000, Kinkhabwala2002}.
This results in ambiguity between the diagnostic of the electron density and that of the photo-exciting flux, which is related to the distance from the source.

\citet{Kaastra2004} calculated a photo-ionization model for Be-like O$^{+4}$ and included PE. They considered two K$\alpha$ (1s-2p) absorption lines from the meta-stable level 1s$^2$2s2p$_{1/2} (J=0)$  at $\sim22.45$\,\AA, which is sensitive to density around $n_e \cong 10^{10}$\,cm$^{-3}$. \citet{Mauche2003} showed that the 2p-3d transition from the  1s$^2$2s$^2$2p$_{3/2} (J=3/2)$ meta-stable level in B-like Fe$^{+21}$ at 11.92\,\AA\, can be a useful $n_e$ or PE diagnostic. 
\citet{Miller2008} used the Fe 11.92\AA\ absorption line to measure a density of $n_e = 5^{+1.3}_{-1.0}\times10^{13}$\,cm$^{-3}$ in the 2005 outburst of the X-ray binary \gro\;and claimed that PE is negligible.
A comprehensive work on the population of meta-stable levels of L-shell ions in astrophysically abundant elements between C and Fe was reported by \citet{Mao2017}, where diagnostics were calculated for densities up to $10^{14}$\,cm$^{-3}$.

In steady-state photo-ionized plasmas, the electron density also determines the ionization of the plasma, since electron recombination  balances photo-ionization. The ionization level of the gas is quantified through the ionization parameter $\xi$  \citep{Tarter1969}, usually expressed in cgs units of erg\,cm\,s$^{-1}$
\begin{equation}
    \label{eq:xi}
   \xi=\frac{L}{n_e r^2}
\end{equation}

\noindent where $L$ (erg\,s$^{-1}$) is the luminosity of the source, and $r$ (cm) is the distance from the ionizing source. If $\xi$ is estimated from the observed ion and $L$ is measured, 
the density diagnostic reveals $r$. Indeed, from $\xi=10^5$ \citet{Miller2008} inferred an outflow distance of $r = 10^{9}$\,cm, which led to a conclusion that thermal driving is not viable for the \gro\;outflow. 
In their level population model they neglected PE and included only electron collisional excitation (CE). 
A recent calculation by \cite{Tomaru2023} included PE, and a lower measured population of the meta-stable level. 
Accordingly, they obtained a lower value of $n_e=2.8 \times 10^{12}$\,cm$^{-3}$ and larger distance of $\cong10^{11}$\,cm, which makes thermal driving also possible.

Since both photo-ionization (PI) and PE processes are driven by the source flux, it is natural to expect that if PI is the dominant ionization process, PE will be an important excitation process. In photo-ionized plasmas, for a given $\xi$, a high $n_e$ implies a lower $r$ and therefore an increased role of PE, and vice versa. In other words, the effects of CE ($n_e$) and PE ($r$) on the population of meta-stable levels must go hand in hand. 
This was shown by \citet{Peretz2019} who measured the distance from the nucleus of NGC\,4051 using the He-like triplets of O and N. Indeed, CE and PE models for the same $\xi$ resulted there in similar distances of $r\cong10^{15}$\,cm, differing by only a factor of 2. 

The goal of the present work is to determine under which physical circumstances the contributions of PE and CE are important, and whether generally both of them need to be considered. We first use an analytic hydrogenic approximation over a wide range of plasma conditions and photo-ionizing sources (Sec. \ref{sec:method}). 
We then use atomic data for Fe$^{+21}$ to build a collisional-radiative model with PE to demonstrate this is the case for the 2005 \gro\ outburst (Sec. \ref{sec:res}). 
We present our conclusions in Sec.\,\ref{sec:conclusion}.


\section{Method} \label{sec:method}
\subsection{Energy Level Kinetics} \label{subsec:energy}
 The observed emission and absorption spectral lines in plasmas depend on the populations of the corresponding energy levels $i$ and $j$ ($i<j$). Therefore, for spectral modeling a detailed computation of their number densities $n_i,n_j$ is needed. The total emission line intensity (ph\,s$^{-1}$cm$^{-3}$) at the source is:
 
 \begin{equation}
    \label{eq:emiss}
   I_{ji}=n_jA_{ji}
\end{equation}

\noindent where $A_{ij}\,(s^{-1})$ is the Einstein spontaneous emission coefficient. The absorption line intensity is:

\begin{equation}
    \label{eq:absorb}
   I_{ij}=I_{ij,0}e^{-\int n_i\sigma_{ij}dl}
\end{equation}

\noindent where $I_{ij,0}$ is the incident intensity around the line, $\sigma_{ij}$\;(cm${^2}$) is the PE cross section, and the integral is over the line of sight. In the general case, a proper computation of these densities requires considering both CE and PE.
 
In hot dense plasmas, the electron impact collisions, which depend on the electron density $n_e$ and temperature $T_e$, are predominantly responsible for ionizing and exciting the ions. The CE rate coefficient ($\rm{cm}^3\,\rm{s}^{-1}$)  from level $i$ to $j$ can be approximated by the \citet{van1962} formula:

\begin{equation}
    \label{eq:coll_exc}
   S_{ij}(T_e)=1.7\times10^{-3}f_{ij}\overline{g}E_{ij_{[eV]}}^{-1}T^{-1/2}_{e[K]}e^{-\frac{E_{ij}}{kT_e}}
\end{equation}

\noindent where $E_{ij}=E_j-E_i$ is the line energy in eV, and $f_{ij}$ is the oscillator strength of the transition; $\overline{g}$ is the averaged excitation Gaunt factor which is of the order of unity. 
From the detailed balance argument in thermodynamic equilibrium, the rate coefficient $S'_{ji}$ of the opposite process, collisional de-excitation from $j$ to $i$ is related to $S_{ij}$ by:
\begin{equation}
    \label{eq:coll_deexc}
   S'_{ji}=\frac{g_i}{g_j}e^{\frac{E_{ij}}{kT_e}}S_{ij}
\end{equation}

\noindent where $g_i$ and $g_j$ are the statistical weights of levels $i, j$, respectively.

In photo-ionized plasma, an external radiation source dominates the ionization and excitation processes. 
The PE rate (s$^{-1}$) from level $i$ to $j$ is given by:
\begin{equation}
    \label{eq:ph_exc}
   R_{ij}^{PE}=\int F_{\nu}(\nu)\,\sigma_{ij}^{PE}d\nu
\end{equation}

\noindent where $F_{\nu}\:(\rm{ph\,s^{-1}cm}^{-2}\rm{Hz^{-1}})$ is the photon flux density, and the cross section $\sigma_{ij}^{PE}$ (cm$^{2}$) is given by:

\begin{equation}
    \label{eq:sigma_ph}
   \sigma_{ij}^{PE}=\frac{\pi e^2}{m_e c}f_{ij}\phi(\nu)
\end{equation}

\noindent where $e,m_e$ are the electron charge and mass, respectively, and $\phi(\nu)$ is a normalized narrow line profile around the transition frequency. $F_{\nu}(\nu)$ can be rewritten as:
\begin{equation}
    \label{eq:flux_dens}
   F_{\nu}(\nu)=hF_E(E)=h\frac{L_E(E)}{4\pi r^2}=\frac{hLf_E(E)}{4 \pi r^2 E}=\frac{n_e \xi h f_E(E)}{4 \pi E}
\end{equation}
where $h$ is the Planck constant, $F_E$ is given in units of $\rm{ph\,s^{-1}cm}^{-2}\rm{keV^{-1}}$, $L_E\:(\rm{ph\,s^{-1}}\rm{keV^{-1}})$ is the photon luminosity density, and $f_E(E)$ is the normalized spectral energy distribution (SED), usually taken from 1-1000\,Ry. In Eq.\,\ref{eq:flux_dens} we used the expression for $\xi$ of Eq.\,\ref{eq:xi} with the energy luminosity $L=\int EL_EdE$.
By substituting Eqs.\,\ref{eq:sigma_ph} and \ref{eq:flux_dens} into Eq.\,\ref{eq:ph_exc} one gets:
\begin{equation}
    \label{eq:ph_exc2}
   R_{ij}^{PE}= n_e \frac{h e^2 \xi f_{ij} f_E(E_{ij}) }{4m_e c E_{ij}}
\end{equation}

\subsection{Two-level approximation} \label{subsec:analytic}  
In this section, we compare the relative roles of CE and PE in populating level $j$ from level $i$, for different ionizing sources and absorbing plasma conditions. Before solving the full collisional-radiative model with PE, which takes into account all populating and depopulating processes of all levels, we focus on a two-level system. 
The ratio of the CE and PE rates (Eqs.\,\ref{eq:coll_exc}, \ref{eq:ph_exc2}) for a specific $i$ to $j$ transition is:

\begin{equation}
    \label{eq:pecoll_ratio}
   \frac{R_{ij}^{PE}}{n_e S_{ij}}=\frac{h e^2 \xi f_E(E_{ij})f_{ij}}{4S_{ij} E_{ij} m_e c }=8.7\frac{\xi}{10^3} \left(  \frac{T_e}{10^6 K} \right)^{1/2} f_E(kT_e)
\end{equation}

\noindent where we assumed $E_{ij}= kT_e$ and $\mathbf{\overline{g}=1}$. 
The above ratio is computed here for two types of SEDs, a power-law (PL) with photon index $1.5<\Gamma<2.5$, typical of AGNs, and a multiple-$T$ black body disc spectrum with maximal source temperature $0.3\,{\rm keV}<kT_s<3$\,keV, typical of XRBs. 
We compute this ratio for two different $T_e$ values and for $\xi=10^3$, which is  the ionization parameter of peak formation of the Fe$^{+21}$ ion \citep[][Sec.\,\ref{sec:res}]{Kallman2009}, but the scaling with $\xi$ is trivial.

The ratios computed from Eq.\,\ref{eq:pecoll_ratio} are presented in Table \ref{tb:pece_values_ratio}. The disc spectra are computed with the {\it diskbb} model \citep{Mitsuda1984} in Xspec \citep{Arnaud1996}.
Note that since $f_E \propto E^{-\Gamma+1}$, the special case of $\Gamma=1.5$ eliminates the 
$T_e^{1/2}$ dependence.
The tabulated ratios show that the PE/CE rate ratios range between $0.3 - 20$.
This demonstrates that both CE and PE can be significant and need to be considered in most cases. Specifically, the \gro\;outburst, which is discussed in Sec.\,\ref{sec:res}, has a {\it diskbb} spectrum with $kT_s=1.3$\,keV \citep{Miller2008}, where the ratio of Eq.\,\ref{eq:pecoll_ratio} is of the order of unity (red text in Table \ref{tb:pece_values_ratio}). 

\begin{deluxetable}{cccccccccc}[h]
\tabletypesize{\scriptsize}
\tablewidth{0pt}
\label{tb:pece_values_ratio}
\tablecaption{PE/CE rate ratio for power-law and {\it diskbb} spectra (Eq.\ref{eq:pecoll_ratio}) with $\xi=10^3$. 
The $kT_s=1.3$\,keV case (in red) refers to \gro\ \citep{Miller2008}.}
\tablehead{
\colhead{\hspace{0.02\textwidth}}&
\colhead{\vline}&
\colhead{\hspace{0.02\textwidth}}&
\colhead{\hspace{0.02\textwidth}Power-law}&
\colhead{\hspace{0.02\textwidth}}&
\colhead{\vline}&
\colhead{\hspace{0.02\textwidth}}&
\colhead{\hspace{0.1\textwidth}{\it diskbb}}&
\colhead{\hspace{0.02\textwidth}}&
\colhead{\hspace{0.02\textwidth}}
}
\startdata
\hspace{0.02\textwidth}log($T_e$)\,[K]&
\vline&
\hspace{0.02\textwidth}$\Gamma$=1.5&
\hspace{0.02\textwidth}$\Gamma$=2&
\hspace{0.02\textwidth}$\Gamma$=2.5&
\vline&
\hspace{0.02\textwidth}$kT_s$=0.3\,keV&
\hspace{0.02\textwidth}$kT_s$=1\,keV&
\hspace{0.02\textwidth}$kT_s$=1.3\,keV&
\hspace{0.02\textwidth}$kT_s$=3\,keV\\
\hline
\hspace{0.02\textwidth}6&
\vline&
\hspace{0.02\textwidth}4.15&
\hspace{0.02\textwidth}14.62&
\hspace{0.02\textwidth}20.71&
\vline&
\hspace{0.02\textwidth}7.36&
\hspace{0.02\textwidth}1.52&
\hspace{0.02\textwidth}\textbf{\textcolor{red}{$\mathbf{1.07}$}}&
\hspace{0.02\textwidth}0.38  \\
\hspace{0.02\textwidth}7&
\vline&
\hspace{0.02\textwidth}4.15&
\hspace{0.02\textwidth}4.62&
\hspace{0.02\textwidth}2.07&
\vline&
\hspace{0.02\textwidth}15.37&
\hspace{0.02\textwidth}8.48&
\hspace{0.02\textwidth}\textbf{\textcolor{red}{$\mathbf{6.41}$}}&
\hspace{0.02\textwidth}2.55 \\
\enddata
\end{deluxetable}
\vspace{-1.5cm}

Another process that can populate excited levels is radiative recombination (RR), in which an electron is captured into a bound state and a photon is emitted.
The rate coefficient for RR to the ground level $\alpha^{RR}\:(\rm{cm}^3\,\rm{s}^{-1})$ is given by\,\citet{Cillie1932}:

\begin{equation}
    \label{eq:rr_rate}
    \alpha^{RR}=3.2\times10^{-6}\frac{e^{E_{ion} / kT_e}}{\mathbf{n^3} T_e^{3/2}}E_1 \left( \frac{E_{ion}}{kT_e} \right)
\end{equation}

\noindent where $n$ is the principal quantum number, $E_{ion}$ is the ionization energy, and $E_1$ is the exponential integral:

\begin{equation}
    \label{eq:E_1}
    E_1(x)=\int_x^\infty \frac{e^{-t}}{t} dt
\end{equation}

The ratio between the PE and RR rates is (Eqs.\,\ref{eq:ph_exc2}, \ref{eq:rr_rate}):
\begin{equation}
    \label{eq:perr_ratio}
    \frac{R_{ij}^{PE}}{n_e \alpha^{RR}}=1.7\times10^{-6} \frac{\xi f_E(E_{ij})f_{ij} T_e^{3/2}}{E_{ij} E_1\left(\frac{E_{ion}}{kT_e} \right)} e^{-E_{ion}/kT_e} = 3.2\times10^6 \frac{\xi}{10^3} \left(  \frac{T_e}{10^6 K} \right)^{1/2} f_E(kT_e)
\end{equation}

\noindent In the last equality we assumed $E_{ij}= E_{ion} = kT_e$ and $f_{ij} = 0.1$. The value of a few millions on the right hand side indicates that recombination is likely negligible for all source conditions. 
The ratio, computed for two different spectral shapes and two temperatures (similar to Table \ref{tb:pece_values_ratio}) is presented in Table\,\ref{tb:perr_values_ratio}. 



\begin{deluxetable}{cccccccccc}[h]
\tabletypesize{\scriptsize}
\tablewidth{0pt}
\label{tb:perr_values_ratio}
\tablecaption{PE/RR rate ratio \textbf{in millions ($\mathbf{\times10^6}$)}, for power-law and {\it diskbb} spectra (Eq.\ref{eq:perr_ratio}) with $\xi=10^3$. 
The $kT_s=1.3$\,keV case (in red) refers to \gro\ \citep{Miller2008}}.
\tablehead{
\colhead{\hspace{0.02\textwidth}}&
\colhead{\vline}&
\colhead{\hspace{0.02\textwidth}}&
\colhead{\hspace{0.02\textwidth}Power-law}&
\colhead{\hspace{0.02\textwidth}}&
\colhead{\vline}&
\colhead{\hspace{0.02\textwidth}}&
\colhead{\hspace{0.1\textwidth}{\it diskbb}}&
\colhead{\hspace{0.02\textwidth}}&
\colhead{\hspace{0.02\textwidth}}
}
\startdata
\hspace{0.02\textwidth}log($T_e$)\,[K]&
\vline&
\hspace{0.02\textwidth}$\Gamma$=1.5&
\hspace{0.02\textwidth}$\Gamma$=2&
\hspace{0.02\textwidth}$\Gamma$=2.5&
\vline&
\hspace{0.02\textwidth}$kT_s$=0.3\,keV&
\hspace{0.02\textwidth}$kT_s$=1\,keV&
\hspace{0.02\textwidth}$kT_s$=1.3\,keV&
\hspace{0.02\textwidth}$kT_s$=3\,keV\\
\hline
\hspace{0.02\textwidth}6&
\vline&
\hspace{0.02\textwidth}1.55&
\hspace{0.02\textwidth}5.47&
\hspace{0.02\textwidth}7.75&
\vline&
\hspace{0.02\textwidth}2.75&
\hspace{0.02\textwidth}0.57&
\hspace{0.02\textwidth}\textbf{\textcolor{red}{$\mathbf{0.40}$}}&
\hspace{0.02\textwidth}0.14  \\
\hspace{0.02\textwidth}7&
\vline&
\hspace{0.02\textwidth}1.55&
\hspace{0.02\textwidth}1.72&
\hspace{0.02\textwidth}0.77&
\vline&
\hspace{0.02\textwidth}5.75&
\hspace{0.02\textwidth}3.17&
\hspace{0.02\textwidth}\textbf{\textcolor{red}{$\mathbf{2.40}$}}&
\hspace{0.02\textwidth}0.95 \\
\enddata
\end{deluxetable}
\vspace{-1.5cm}

Both PE/CE and PE/RR ratios decrease with increasing temperature for a PL spectrum, and increase with temperature for a {\it diskbb} spectrum. This is due to the energy dependence of $f_E$, which can decrease faster or slower with $E$, compared to $T_e^{1/2}$ (Eq.\,\ref{eq:pecoll_ratio}, Eq.\,\ref{eq:perr_ratio}). 
This two-level model indicates that in all cases where $kT_e\sim E_{ion}$, including \gro\;(red text in Table \ref{tb:perr_values_ratio}), the PE/RR ratio is $\sim10^5 - 10^6$. Hence, RR can be safely neglected, while PE must be considered.

We can also consider a lower temperature of $T_e=10^5$\,K, for which $\xi$ will also be lower $\xi\approx10^{1.5}$ \citep[{See Fig\,.6 in}][]{Kallman2004}. For a PL spectrum, the PE/CE rate ratios are between 0.1-4 and the PE/RR ratios are (0.05-1.4)$\times10^6$. 
This demonstrates that at lower temperatures as well, CE and PE are both important, while RR is not. 
$T_e=10^5$\,K is not considered here for {\it diskbb} spectra, as typical X-ray binaries have much higher radiation temperatures $kT_s > 0.3$\,keV.

Photo-ionization (PI) transitions can also (populate or) depopulate the excited levels of the ion. The PI rate from level $i$ is given in a similar fashion to that of the PE rate (Eq.\ref{eq:ph_exc}):

\begin{equation}
    \label{eq:ph_ion}
   R_{i}^{PI}=\int F_{\nu}(\nu)\,\sigma_{i}^{PI}d\nu
\end{equation}
\noindent $\sigma_{i}^{PI}$(cm$^2$) is the PI cross section, which is given by \citep{Karzas1961}:

\begin{equation}
    \label{eq:sigma_pi}
   \sigma_{i}^{PI}= \left( \frac{64\pi n g}{3\sqrt{3}} \right) \alpha a_0^2 \left( \frac{E_{ion}}{E} \right)^3
\end{equation}

\noindent where $g$ is the Gaunt factor which is of the order of unity, $a_0=\hbar^2/m_e e^2$ is the Bohr radius, and $\alpha=e^2/\hbar c$ is the fine structure constant.
To demonstrate the difference in orders of magnitude between PE and PI, we evaluate the rates at $E=E_{ion}=1$\,keV (Eqs.\,\ref{eq:ph_exc}, \ref{eq:sigma_ph}, \ref{eq:flux_dens}, \ref{eq:ph_ion}, \ref{eq:sigma_pi}):

\begin{equation}
    \label{eq:pepi_ratio}
    \frac{R_{ij}^{PE}}{R_i^{PI}}(E=E_{ion}=1\,\mathrm{keV})=\frac{0.002\times f_E(1\,keV)}{8\times10^{-18}}=2.5\times10^{13}
\end{equation}

\noindent This result shows that the PI population of excited levels is insignificant compared to PE. This is not surprising given that PE is a resonant process. The PI rates are computed fully in our numerical analysis in Sec.\,\ref{sec:res}, but there too they are negligible.

\subsection{Collisional radiative model}
\label{subsec:Coll_rad}
In order to solve the populations of all levels of a given ion, we refer to a collisional radiative model with PE. The level $i$ population $n_i$ is solved from a set of linear rate equations.
\begin{equation}
    \label{eq:rate_eq}
   \frac{dn_i}{dt}=0=\sum_{j<i} \left( n_j n_e S_{ji}+n_j R_{ji}^{PE} \right) +\sum_{k>i} \left( n_k A_{ki}+n_k n_e S'_{ki} \right) -\sum_{j<i} \left( n_i A_{ij}+n_i n_e S'_{ij} \right) -\sum_{k>i} \left( n_i n_e S_{ik}+n_i R_{ik}^{PE} \right)-\mathbf{n_i R^{PI}_i}
\end{equation}

\noindent The first and second terms represent the populating processes of level $i$ from lower and higher levels. The third and fourth terms, represent the depopulating processes of level $i$ to lower and higher levels, the last term is the depopulating PI.
In a steady state, i.e. $dn_i/dt=0$, Eq.\,\ref{eq:rate_eq} reduces to a set of linear algebraic equations, which is solved to get the set of population densities \{$n_i$\}. 
These results are presented in the following.

\section{Results for F\MakeLowercase{e}$^{+21}$} 
\label{sec:res}
In this section, we consider the special case of B-like Fe$^{+21}$, which was debated in the context of the launching mechanism of the \gro\ outflow observed by {\it Chandra}/HETGS in 2005 \citep{Miller2006,Miller2008,Netzer2006,Tomaru2023}. 
The atomic structure of Fe$^{+21}$ and the 11.92\,\AA/11.77\,\AA\ line ratio density diagnostics were discussed in detail by \citet{Mauche2003}. 
References to previous uses of these lines in solar and laboratory plasmas can be found there, as well as excellent physical insights into the origin of the density sensitivity of these lines.
Here, it is sufficient to understand that at low densities (and no PE) practically all ions are in the ground level $2s^22p_{1/2} (n_1)$ from which the 11.77\,\AA\ originates. 
Only at higher densities (or significant PE) does the first excited level $2s^22p_{3/2}$ get populated $(n_2)$, and the 11.92\,\AA\ line arises.
Excitations from the ground level can be divided into $\Delta n = 0$ (2-2, i.e., UV) and $\Delta n = 1$ (2-3, i.e., X-ray) transitions, hence the importance of $T_e$ and the SED.

Our Fe$^{+21}$ model includes 63 energy levels of the following B-like configurations $1s^22l^3$ (10 levels) and $1s^22l^23l'$ (53 levels), with orbital quantum numbers $l=s,p$ and $l'=s,p,d$. 
Consequently, excitations up to level 10 are referred to as UV transitions ($2-2, \Delta n = 0$), while excitations to levels 11-63 are X-ray transitions ($2-3, \Delta n = 1$).

The present analysis focuses on the $n_2/n_1$ population ratio.
As in \citet{Mauche2003}, the atomic structure and coefficients were computed using the relativistic Hebrew University Lawrence Livermore Atomic Code \citep[HULLAC,][]{Bar-Shalom2001}. 

The model (Eq.\,\ref{eq:rate_eq}) provides the $n_2/n_1$ population ratio as a function of density, which is then compared to the $11.92$\AA$/11.77$\AA\:absorption line ratio in the 2005 spectrum of \gro . 
Eq.\,\ref{eq:rate_eq} is solved under three different approximations, in order to demonstrate the effect of each process independently. First, we included only CE terms; second, only PE terms, and lastly, both CE and PE are included.
The spontaneous emission terms are included in all cases.
Since the PE rates scale with the flux $\propto L r^{-2}$ (Eq.\,\ref{eq:ph_exc}), they require a value for $L$. 
Hence, the PE-only populations can be presented as a function of $r$, but will depend on the assumed $L$ value. 
In the CE \& PE model, CE rates depend on $n_e$ (and less so on $T_e$), while PE rates still depend on flux $\propto L r^{-2}$, but these two are tied through $\xi$ (Eq.\,\ref{eq:xi}).
Hence, the CE \& PE populations can be presented as a function of $n_e$ or $r$, but will depend on the assumed $\xi$ (and $T_e$) value. 
In the following, we assume $L=5.7\times10^{37}$\,erg\,s$^{-1}$ \citep{Miller2008} and $\xi = 10^3$ \citep{Kallman2009}, and discuss how the results would change if these values change.
A value of $\xi=10^3$ is appropriate for Fe$^{+21}$ in the \gro\ outburst, as it is where the ion obtains its maximal abundance \citep{Kallman2009}. 
The same value was obtained in \citet{Netzer2006}, while  \citet{Tomaru2023} used $\xi\cong2.5\times 10^3$.
\\

\subsection{Only-CE} \label{sec:onlyce}
\begin{figure}[h]
    \centering
    \includegraphics[width=0.49\textwidth]{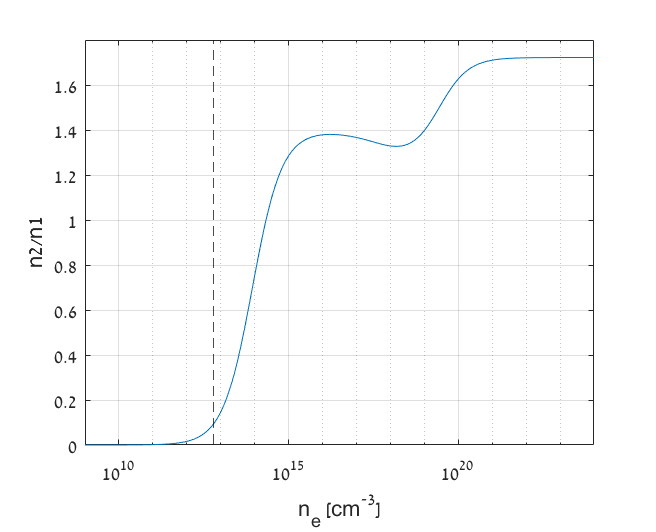}
    \includegraphics[width=0.49\textwidth]{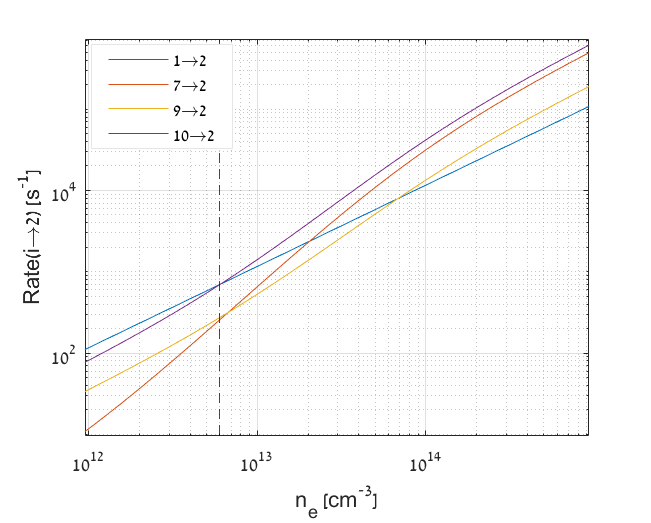}
    
  \caption{\textbf{Left: }$n_2/n_1$ population ratio for Fe$^{+21}$, considering only CE. \textbf{Right:} Four dominant levels that populate the metastable level $n_2$ around the critical density $n_{e,c} = 6\times10^{12}$ cm$^{-3}$ (denoted by the vertical dashed line in both panels).} \label{fig:CE_ratio}
\end{figure}

The $n_2/n_1$ population ratio computed with CE, but not PE, is shown in Fig.\,\ref{fig:CE_ratio} (left panel), assuming $T=10^6$\,K. 
It shows a critical density of $n_{e,c}=6\times10^{12}$ cm$^{-3}$ in which $n_2/n_1$ starts to increase non-linearly. 
At much higher $n_e>10^{21}\,\rm{cm}^{-3}$, $n_2/n_1$ tends to the Boltzmann population ratio, $n_2/n_1=2\,{\rm exp}(-E_{21}/kT)=1.722$, where the gas is in local thermodynamic equilibrium (LTE).
The right hand side panel of Fig.\,\ref{fig:CE_ratio} shows the total rates of the dominant transitions to level 2. At low densities $n_e \ll n_{e,c}$, $n_2$ is populated predominantly by the weak collisional excitation from level 1 to 2 ($2s^22p_{1/2}-2s^22p_{3/2}$), which then decays back to level 1. As $n_e$ approaches $n_{e,c}$, the $2s2p_{3/2}^2$ levels (7, 9, and 10) get increasingly populated, and their total transition rates to level 2 surpass that of 1 to 2, thus dramatically increasing the $n_2/n_1$ ratio. The crossing point of the 1 to 2 and 10 to 2 rates can be used to define the critical density of $n_{e,c}=6\times10^{12}$ cm$^{-3}$.
The increase in the $n_2/n_1$ ratio in the l.h.s. of Fig.\,\ref{fig:CE_ratio} is similar to that found in \citet{Miller2008}, while the results of \citet[][Fig.\,3 therein]{Mao2017} indicate slightly lower densities to reach $n_2/n_1 = 1$.

\subsection{Only-PE} \label{sec:onlype}
\begin{figure}[h]
    \centering
    \includegraphics[width=0.49\textwidth]{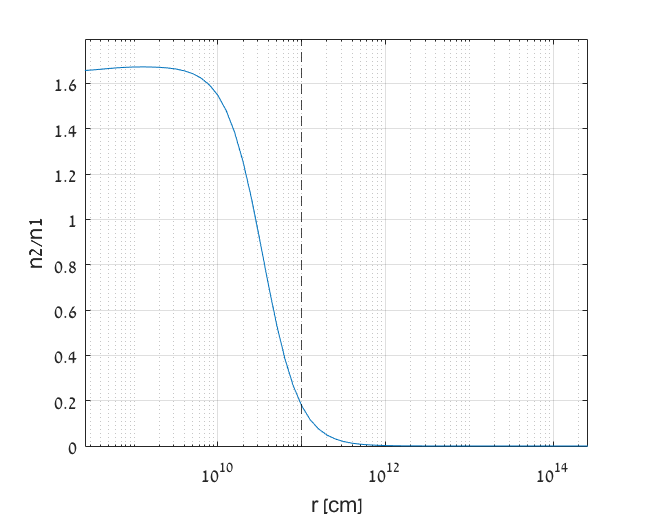}
    \includegraphics[width=0.49\textwidth]{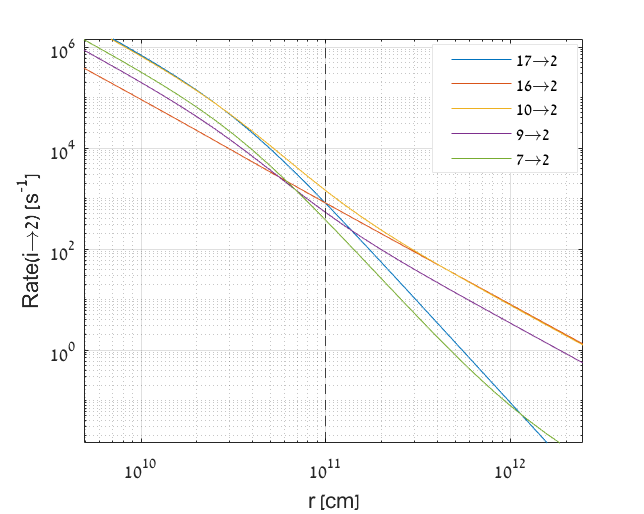}
  \caption{\textbf{Left: }$n_2/n_1$ population ratio of Fe$^{+21}$, considering only PE. \textbf{Right:} Five dominant transitions that populate the metastable level $n_2$ around the critical distance $r_{c}=10^{11}$\,cm (vertical dashed line in both panels).}
   \label{fig:PE_ratio}
\end{figure}

The next case we consider is a model with only the PE terms, and no CE. We used the measured {\it diskbb} spectrum of \gro\ with $kT_s=1.3$\,keV, and a luminosity of $L_X=5\times10^{37}$\,erg\,s$^{-1}$ between 0.65-10\,keV \citep{Miller2008}, which implies a bolometric luminosity of $L=5.7\times10^{37}$\,erg\,s$^{-1}$. The $n_2/n_1$ population ratio is shown in Fig.\,\ref{fig:PE_ratio} as a function of distance from the ionizing source. 



Far from the source ($r \gg r_{c}=10^{11}$\,cm) the exciting flux is low and the effect of PE is weak, thus $n_2 \ll n_1$. As $r$ decreases, the flux ($\propto r^{-2}$) increases, PE of excited levels increases, and radiative cascades increasingly populate level 2. 
Around the critical distance $r_{c}=10^{11}$\,cm, $n_2$ approaches $n_1$.
At closer distances, PE from level 2 becomes significant,
and balances the cascade effect. 
This is seen as the plateau on the low-$r$ side of Fig.\,\ref{fig:PE_ratio}. 
The right hand side panel of Fig.\,\ref{fig:PE_ratio} shows the total transition rates of the dominant levels populating $n_2$ around $r_{c}=10^{11}$\,cm. 
Farther from the source ($r>r_{c}$) level 16 ($2s^23d_{3/2}$), which is photo-excited from level 1 by X-rays (the 11.77\,\AA\ transition) dominates the population of level 2. 
Below the critical distance, $n_2$ increases on the account of $n_1$, thus PE from level 1 to level 16 diminishes, while the rate for PE from level 2 to level 17 ($2s^23d_{5/2}$, which is the X-ray 11.92\,\AA\ transition), and the decay back to level 2, dominate. 
Level 10 has a strong radiative transition to level 2, therefore it populates level 2 at all distances. The crossing point of the 16 to 2 (fed by 1 to 16) and 17 to 2 rates can be used to define the critical distance of $r_c=10^{11}$\,cm,
where level 2 is significantly populated by PE. The plotted distances in Fig.\,\ref{fig:PE_ratio} actually represent fluxes, derived from the assumed $L$. If $L$ is in fact higher than the assumed $5.7\times10^{37}$\,erg\,s$^{-1}$, for example the central source is obscured to us but visible to the outflow, the derived distances would also be higher ($r \propto \sqrt L$).

\subsection{Final Results} \label{sec:cepe}

\begin{figure}[h]
    \centering
    \includegraphics[width=0.86\textwidth]{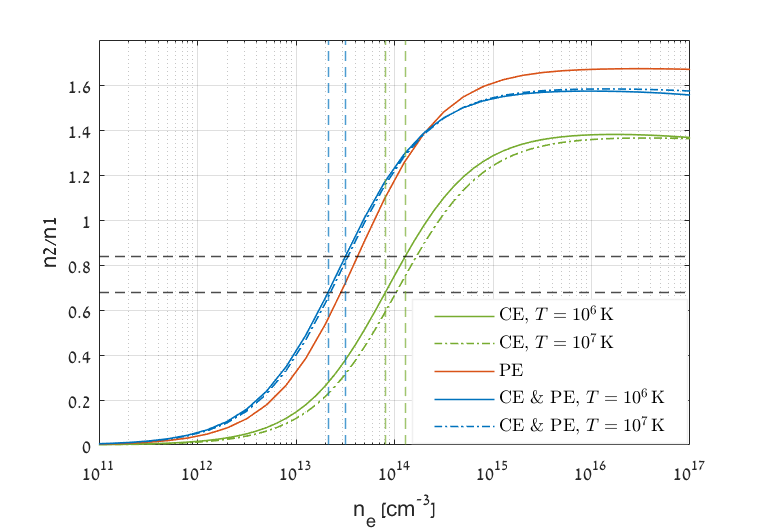}
  \caption{The $n_2/n_1$ population ratio of Fe$^{+21}$ for three models of only-CE, only-PE, and CE \& PE. CE was computed for $T=10^{6},10^{7}$\,K. The horizontal black dashed lines are the presently measured upper and lower limits of $n_2/n_1$. Vertical dashed lines are the inferred upper and lower limits for $n_e$. PE can be seen to have the dominant contribution and neglecting it results in an overestimate of $n_e$ by a factor of 4.}
   \label{fig:CEPE_ratio}
\end{figure}

Our final results for $n_2/n_1$ when including both CE and PE are presented in Fig.\,\ref{fig:CEPE_ratio}, and compared to only-CE and only-PE models at two temperatures of $T=10^{6},10^{7}$\,K. 
The difference between $T=10^{6}$\,K and $T=10^{7}$\,K is statistically insignificant. The figure shows that neglecting PE results in an $n_e$ value that is higher by a factor of 4, and thus a distance smaller by a (square root) factor of 2. The CE \& PE model for Fe$^{+21}$ assumes $\xi=10^3$; A higher value will move the CE \&  PE curve to lower densities since the PE rates will be higher (c.f., Eq.\,\ref{eq:ph_exc2}). 

\begin{figure}[h]
    \centering
    \includegraphics[width=0.43\textwidth]{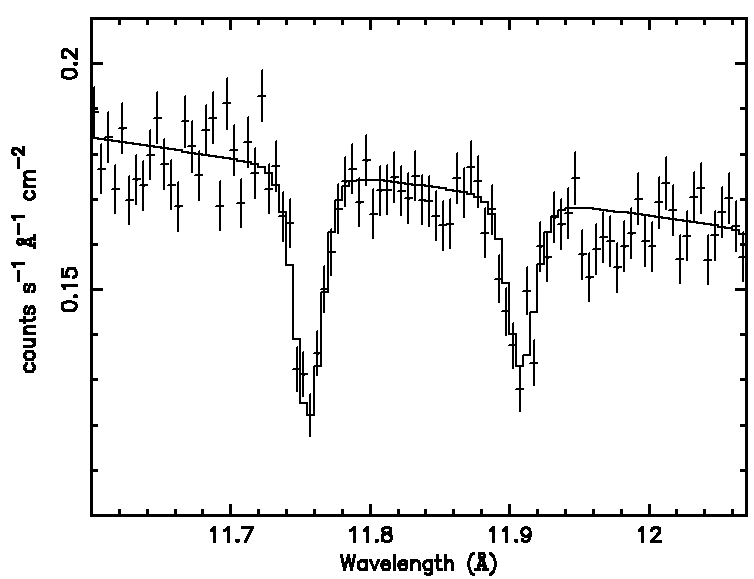}
    \includegraphics[width=0.50\textwidth]{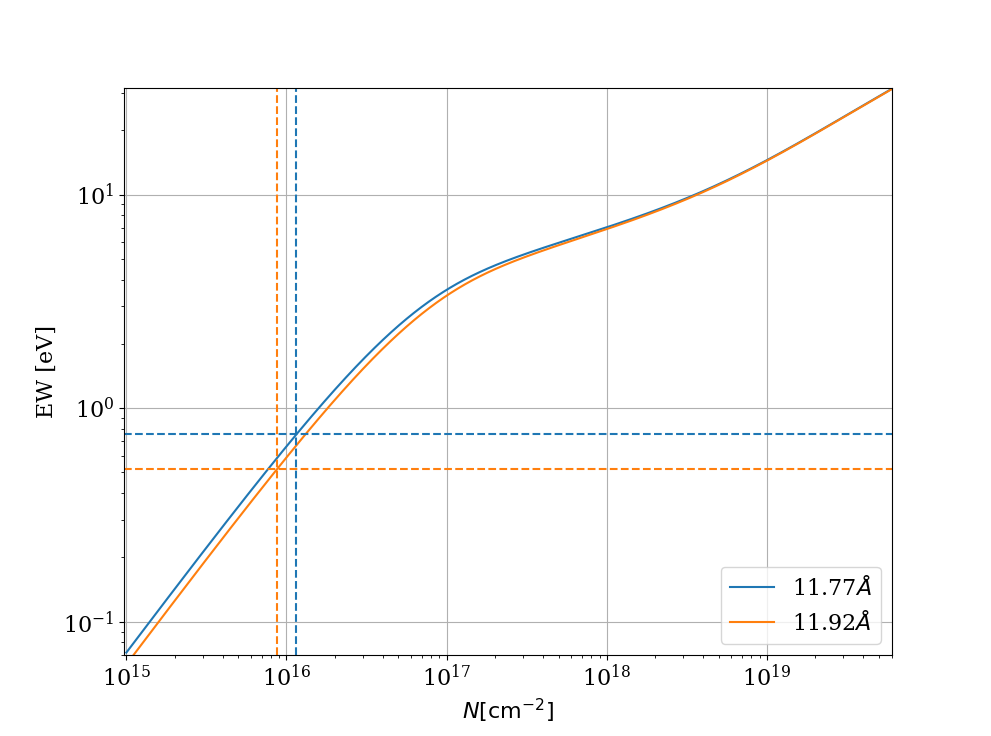}
  \caption{\textbf{Left: }Voigt profile fits to the absorption lines observed in the \emph{Chandra}/HETG MEG spectrum of \gro, between 11.6-12.06\,\AA.
  \textbf{Right: }Theoretical curves of growth for the 11.77\,\AA\ (1 to 16) and 11.92\,\AA\ (2 to 17) absorption lines, with their similar  
oscillator strengths of $f_{11.77\rm{\mathring{A}}}=0.66$ and $f_{11.92\rm{\mathring{A}}}=0.58$ (from  HULLAC). 
Horizontal and vertical lines mark the measured EWs and the inferred column densities for each line.
}
   \label{fig:eqw}
\end{figure}

We remeasured the equivalent widths (EWs) of  the 11.77\,\AA\;and 11.92\,\AA\ lines by fitting Voigt profiles (Fig.\,\ref{fig:eqw}, left panel) and obtained values of ($8.4\pm0.5$)\,m\AA\ and ($6.0\pm0.5$)\,m\AA, respectively, providing a ratio of $0.71 \pm 0.07\;(1\sigma$). $n_1$ and $n_2$ were obtained by comparison of the EWs to a theoretical curve of growth (Fig.\,\ref{fig:eqw}, right panel). 
This results in $n_2/n_1=0.76 \pm 0.08$, which is consistent with $n_2/n_1=0.71^{+0.05}_{-0.09}$ in \citet{Miller2008}, but not with $n_2/n_1\cong 0.2 \pm 0.1$ in \citet{Tomaru2023}, who presumably obtained the EWs of the lines from a global fit.
Plotting the $n_2/n_1$ confidence region on the theoretical density curve in Fig.\,\ref{fig:CEPE_ratio}, indicates $n_e=(2.6 \pm 0.5)\times10^{13}\,\rm{cm}^{-3}$.

Table\,\ref{tb:final_res} summarizes the $n_e$ and $r$ values for the three current models: only-CE, only-PE, and CE \& PE; $r$ was calculated using $\xi$. 
The present best estimate for density and distance are $n_e=(2.6 \pm 0.5)\times10^{13}\,\rm{cm}^{-3}$ and $r=(4.4 \pm 0.4)\times10^{10}$\,cm. 
Note how the eventual differences between models are a factor of a few, proving again the comparable roles of CE and PE, as inferred from the simplified the two-level hydrogenic approximation in Sec.\,\ref{subsec:analytic}.
\citet{Miller2008} report a density value of $n_e=(5 \pm 1)\times10^{13}\,\rm{cm}^{-3}$, when PE was neglected, and with a different collisional-radiative model than ours. 
\citet{Tomaru2023} included both CE and PE, but obtained a density of $n_e=(0.28\pm 0.09)\times 10^{13}\,\rm{cm}^{-3}$, because of their low $n_2/n_1$ value (see Fig.\,\ref{fig:CEPE_ratio}). We stress that the different distances obtained by different authors depend appreciably also on their assumed $\xi$ value, as $r\propto \xi^{-1/2}$. 

\begin{deluxetable}{lcc}[h]
\tabletypesize{\footnotesize}
\tablewidth{0pt}
\label{tb:final_res}
\tablecaption{$n_e$ and $r$ for three different models.}
\tablehead{
\colhead{} & \colhead{\hspace{0.08\textwidth}$n_e$ [$10^{13}$\,cm$^{-3}$]}
& \colhead{\hspace{0.08\textwidth}$r$ [$10^{10}$\,cm]}
}
\startdata
{Only CE}     &\hspace{0.08\textwidth}$10.2\pm2.4$ &\hspace{0.08\textwidth}$2.2\pm0.2$ \\
{Only PE}     &\hspace{0.08\textwidth}$3.5\pm0.7$  &\hspace{0.08\textwidth}$3.8\pm0.4$ \\
{CE \& PE}   &\hspace{0.08\textwidth}$2.6\pm0.5$  &\hspace{0.08\textwidth}$4.4\pm0.4$ \\
\enddata
\tablecomments{Computed with $L=5.7\times10^{37}$\,erg\,s$^{-1}$, $T=10^6$\,K, and $\xi=10^3$\,erg cm s$^{-1}$.}
\end{deluxetable}
\vspace{-1.5cm}
\subsection{Critical $\xi$}
\label{subsec:xi}
In the numerical models, when computing the case of only-CE and only-PE separately, the results are independent of $\xi$. 
Only when needing to scale CE vs. PE does the model require an assumption for $\xi$.
On the other hand, one can estimate a critical $\xi_c$ from $n_{e,c}$ and $r_c$ (Figs.\,\ref{fig:CE_ratio}, \ref{fig:PE_ratio}). In the above case of Fe$^{+21}$ we obtain $\xi_c=L/(n_{e,c}r_c^2)=950$, which is almost exactly the value of 10$^3$ obtained from photoionization balance \citep{Kallman2009}, which we used above. 
This provides yet another consistency test for the comparable roles of CE and PE, 
once absorption from excited levels are identified, as also demonstrated in Eq.\,\ref{eq:pecoll_ratio} and in Table \ref{tb:pece_values_ratio}.


\section{Conclusions}
\label{sec:conclusion}
This work considers most generally the collisional and radiative processes that populate excited levels of ions in plasmas.  
Sec.\,\ref{sec:method} presents a semi-analytical two-level hydrogenic approximation for a wide range of plasma temperatures and photon fluxes, which leads to the following conclusions:
\begin{itemize}
\item The CE and PE processes generally have a comparable contribution in populating excited levels. The PE/CE rate coefficient ratio is in the range of $0.3-20$.
\item The RR contribution is significantly lower. The PE/RR ratio is in the range of $\sim10^{5}-10^{6}$, hence RR can be safely neglected in the analysis.
\end{itemize}

A collisional radiative model with PE is employed for computing the population ratios of B-like Fe$^{+21}$, which has been extensively studied for the case of the \gro\;2005 outburst. Our conclusions are:

\begin{itemize}
\item At low densities ($n_e \ll n_{e,c}$), and at large distances ($r \gg r_c$), the ions remain predominantly in the ground level and transitions between excited levels are negligible. Only when $n_e \ge n_{e,c}$ or $r \le r_c$
do transitions between excited levels become significant, which enables density or distance diagnostics.

\item The absorption transition at 11.92\,\AA\;from the meta-stable level 2 of Fe$^{+21}$ provides powerful density and distance diagnostics. 
The present model shows that PE is dominant and yields $n_e = 2.6\times10^{13}\,\rm{cm}^{-3}$, and $r=4.4\times10^{10}\,\rm{cm}$. Including only CE results in an underestimate of $n_e$ by a factor of $\sim4$, and $r$ by a factor $\sim2$.

\item Both X-ray and UV PE transitions can (indirectly) populate the meta-stable level 2. Therefore, the model results may depend on the broad-band spectrum. In the cases considered here of power-law or {\it diskbb} ionizing spectra, both X-ray and UV excitations contribute.

\item We present a sanity check for a critical $\xi_c$ calculated from $n_{e,c}$ and $r_c$ that confirms a value of $\xi\cong10^3$ for Fe$^{+21}$ in \gro. Interestingly, this means one can estimate $\xi$ when transitions from meta-stable levels are observed without photoionization balance calculations.
\end{itemize}

This work was supported in part by a Center of Excellence of the Israel Science Foundation (grant No. 1937/19).



\bibliography{bibli}{}

\begin{thebibliography}{}
\expandafter\ifx\csname natexlab\endcsname\relax\def\natexlab#1{#1}\fi
\providecommand{\url}[1]{\href{#1}{#1}}
\providecommand{\dodoi}[1]{doi:~\href{http://doi.org/#1}{\nolinkurl{#1}}}
\providecommand{\doeprint}[1]{\href{http://ascl.net/#1}{\nolinkurl{http://ascl.net/#1}}}
\providecommand{\doarXiv}[1]{\href{https://arxiv.org/abs/#1}{\nolinkurl{https://arxiv.org/abs/#1}}}

\bibitem[{{Arnaud}(1996)}]{Arnaud1996}
{Arnaud}, K.~A. 1996, in Astronomical Society of the Pacific Conference Series, Vol. 101, Astronomical Data Analysis Software and Systems V, ed. G.~H. {Jacoby} \& J.~{Barnes}, 17

\bibitem[{{Bar-Shalom} {et~al.}(2001){Bar-Shalom}, {Klapisch}, \& {Oreg}}]{Bar-Shalom2001}
{Bar-Shalom}, A., {Klapisch}, M., \& {Oreg}, J. 2001, \jqsrt, 71, 169, \dodoi{10.1016/S0022-4073(01)00066-8}

\bibitem[{{Cilli{\'e}}(1932)}]{Cillie1932}
{Cilli{\'e}}, G. 1932, \mnras, 92, 820, \dodoi{10.1093/mnras/92.8.820}

\bibitem[{Gabriel \& Jordan(1969)}]{Gabriel1969}
Gabriel, A.~H., \& Jordan, C. 1969, Monthly Notices of the Royal Astronomical Society, 145, 241, \dodoi{10.1093/mnras/145.2.241}

\bibitem[{{Kaastra} {et~al.}(2004){Kaastra}, {Raassen}, {Mewe}, {Arav}, {Behar}, {Costantini}, {Gabel}, {Kriss}, {Proga}, {Sako}, \& {Steenbrugge}}]{Kaastra2004}
{Kaastra}, J.~S., {Raassen}, A.~J.~J., {Mewe}, R., {et~al.} 2004, \aap, 428, 57, \dodoi{10.1051/0004-6361:20041434}

\bibitem[{{Kallman} {et~al.}(2009){Kallman}, {Bautista}, {Goriely}, {Mendoza}, {Miller}, {Palmeri}, {Quinet}, \& {Raymond}}]{Kallman2009}
{Kallman}, T.~R., {Bautista}, M.~A., {Goriely}, S., {et~al.} 2009, \apj, 701, 865, \dodoi{10.1088/0004-637X/701/2/865}

\bibitem[{{Kallman} {et~al.}(2004){Kallman}, {Palmeri}, {Bautista}, {Mendoza}, \& {Krolik}}]{Kallman2004}
{Kallman}, T.~R., {Palmeri}, P., {Bautista}, M.~A., {Mendoza}, C., \& {Krolik}, J.~H. 2004, \apjs, 155, 675, \dodoi{10.1086/424039}

\bibitem[{{Karzas} \& {Latter}(1961)}]{Karzas1961}
{Karzas}, W.~J., \& {Latter}, R. 1961, \apjs, 6, 167, \dodoi{10.1086/190063}

\bibitem[{{Kinkhabwala} {et~al.}(2002){Kinkhabwala}, {Sako}, {Behar}, {Kahn}, {Paerels}, {Brinkman}, {Kaastra}, {Gu}, \& {Liedahl}}]{Kinkhabwala2002}
{Kinkhabwala}, A., {Sako}, M., {Behar}, E., {et~al.} 2002, \apj, 575, 732, \dodoi{10.1086/341482}

\bibitem[{{Mao} {et~al.}(2017){Mao}, {Kaastra}, {Mehdipour}, {Raassen}, {Gu}, \& {Miller}}]{Mao2017}
{Mao}, J., {Kaastra}, J.~S., {Mehdipour}, M., {et~al.} 2017, \aap, 607, A100, \dodoi{10.1051/0004-6361/201731378}

\bibitem[{{Mauche} {et~al.}(2003){Mauche}, {Liedahl}, \& {Fournier}}]{Mauche2003}
{Mauche}, C.~W., {Liedahl}, D.~A., \& {Fournier}, K.~B. 2003, \apjl, 588, L101, \dodoi{10.1086/375684}

\bibitem[{{Miller} {et~al.}(2006){Miller}, {Raymond}, {Fabian}, {Steeghs}, {Homan}, {Reynolds}, {van der Klis}, \& {Wijnands}}]{Miller2006}
{Miller}, J.~M., {Raymond}, J., {Fabian}, A., {et~al.} 2006, \nat, 441, 953, \dodoi{10.1038/nature04912}

\bibitem[{Miller {et~al.}(2008)Miller, Raymond, Reynolds, Fabian, Kallman, \& Homan}]{Miller2008}
Miller, J.~M., Raymond, J., Reynolds, C.~S., {et~al.} 2008, The Astrophysical Journal, 680, 1359, \dodoi{10.1086/588521}

\bibitem[{{Mitsuda} {et~al.}(1984){Mitsuda}, {Inoue}, {Koyama}, {Makishima}, {Matsuoka}, {Ogawara}, {Shibazaki}, {Suzuki}, {Tanaka}, \& {Hirano}}]{Mitsuda1984}
{Mitsuda}, K., {Inoue}, H., {Koyama}, K., {et~al.} 1984, \pasj, 36, 741

\bibitem[{{Netzer}(2006)}]{Netzer2006}
{Netzer}, H. 2006, \apjl, 652, L117, \dodoi{10.1086/510067}

\bibitem[{{Peretz} {et~al.}(2019){Peretz}, {Miller}, \& {Behar}}]{Peretz2019}
{Peretz}, U., {Miller}, J.~M., \& {Behar}, E. 2019, \apj, 879, 102, \dodoi{10.3847/1538-4357/ab23ef}

\bibitem[{{Porquet} \& {Dubau}(2000)}]{Porquet2000}
{Porquet}, D., \& {Dubau}, J. 2000, \aaps, 143, 495, \dodoi{10.1051/aas:2000192}

\bibitem[{{Sako} {et~al.}(2000){Sako}, {Kahn}, {Paerels}, \& {Liedahl}}]{Sako2000}
{Sako}, M., {Kahn}, S.~M., {Paerels}, F., \& {Liedahl}, D.~A. 2000, \apjl, 543, L115, \dodoi{10.1086/317282}

\bibitem[{{Tarter} {et~al.}(1969){Tarter}, {Tucker}, \& {Salpeter}}]{Tarter1969}
{Tarter}, C.~B., {Tucker}, W.~H., \& {Salpeter}, E.~E. 1969, \apj, 156, 943, \dodoi{10.1086/150026}

\bibitem[{{Tomaru} {et~al.}(2023){Tomaru}, {Done}, \& {Mao}}]{Tomaru2023}
{Tomaru}, R., {Done}, C., \& {Mao}, J. 2023, \mnras, 518, 1789, \dodoi{10.1093/mnras/stac3210}

\bibitem[{{van Regemorter}(1962)}]{van1962}
{van Regemorter}, H. 1962, \apj, 136, 906, \dodoi{10.1086/147445}

\end{thebibliography}
\bibliographystyle{aasjournal}



\end{document}